\documentclass[a4paper,twocolumn,showpacs,superscriptaddress,prb]{revtex4}

\usepackage{graphicx}
\usepackage[english]{babel}
\usepackage{xspace}
\usepackage[normalem]{ulem}
\usepackage{color}
\usepackage{amsmath,amssymb}

\begin{document}

\title{Direct observation of band bending in topological insulator Bi$_2$Se$_3$}

\author {C.E. ViolBarbosa}
\affiliation{Max Planck Institute for Chemical Physics of Solids,
             N{\"o}thnitzer Str. 40, 01187 Dresden, Germany}

\author {Chandra Shekhar}
\affiliation{Max Planck Institute for Chemical Physics of Solids,
             N{\"o}thnitzer Str. 40, 01187 Dresden, Germany}

\author {Binghai Yan}
\affiliation{Max Planck Institute for Chemical Physics of Solids,
             N{\"o}thnitzer Str. 40, 01187 Dresden, Germany}
\affiliation{Max Planck Institute for the Physics of Complex Systems,N{\"o}thnitzer Str. 38, 01187 Dresden, Germany}

\author {S. Ouardi}
\affiliation{Max Planck Institute for Chemical Physics of Solids,
             N{\"o}thnitzer Str. 40, 01187 Dresden, Germany}

\author {G.H. Fecher}
\affiliation{Max Planck Institute for Chemical Physics of Solids,
             N{\"o}thnitzer Str. 40, 01187 Dresden, Germany}

\author {C. Felser}

\affiliation{Max Planck Institute for Chemical Physics of Solids,
             N{\"o}thnitzer Str. 40, 01187 Dresden, Germany}

\email{carlos.barbosa@cpfs.mpg.de}
\date{received 18$^{th}$ December 2013}
\begin{abstract}

 The surface band bending tunes considerably the surface band structures and 
transport properties in topological insulators. We present a 
direct measurement of the band bending on the Bi$_2$Se$_3$  by using the 
bulk sensitive angular-resolved hard x-ray photospectroscopy (HAXPES). We 
tracked the depth dependence of the energy shift of Bi and Se core states. We 
estimate that the band bending extends up to about 20 nm  into the bulk with an 
amplitude of 0.23--0.26 eV, consistent with profiles previously deduced from the binding energies
of surface states in this material.

\end{abstract}

 \maketitle

Topological insulators (TIs), a new quantum state, are characterized by robust 
metallic surface states inside the bulk energy gap, which are due to the 
topology of bulk band structures. ~\cite{qi2010,moore2010,hasan2010,qi2011RMP} A 
large amount of efforts were devoted to observe the topological surface states 
of many TI materials~\cite[and references therein]{yan2012rpp} by 
surface-sensitive experiments. Specially, Bi$_2$Se$_3$ is one of the most 
extensively studied TI materials because of its simple Dirac-type surface states 
and large bulk gap.~\cite{zhang2009,xia2009,chen2010b}

Surface band bending (SBB) effects of Bi$_2$Se$_3$ has been commonly observed 
in angle-resolved photoemission spectra~\cite{hsieh2009,chen2010b} (ARPES) and 
transport measurements~\cite{kong2011,Checkelsky2011}.  The SBB is usually 
caused by surface degrading in ambient environment and surface 
doping~\cite{bianchi2010,King2011,benia2011,hsieh2011}, with a downshift of the 
surface Dirac point, indicating an electron-doped surface~\cite{Koleini2013}.  
SBB induces a quantum confinement effect~\cite{Bahramy2012} and modifies the 
surface and bulk bands dramatically. A clear feature of SBB in Bi$_2$Se$_3$ is a pair of 
Rashba-splitting bands above the Dirac cone.
  In transport experiments, SBB is also supposed to affect the measurement in a 
considerable way by directly tuning the bulk and surface charge carrier 
densities.
  ~\cite{Steinberg2010,Checkelsky2011,Xiu2011,Kim2012} So far, this surface band 
bending has only been deduced~\cite{benia2011,King2011} from Rashba-splitting of 
the conduction bands measured by ARPES, that is mainly sensitive to several 
surface atomic layers, although SBB is predicted to extends in an order of 10 nm 
distance from the surface into the bulk. A direct measurement of SBB from the 
surface into the bulk region is yet to be performed.

In this Letter, we reported the direct observation of SBB on the Bi$_2$Se$_3$ 
surface by HArd X-ray PhotoElectron Spesctroscopy (HAXPES), a bulk sensitive 
method. The hard x-ray excitation ($\sim$8keV) produces photoelectrons with high 
kinetic energy and consequently high inelastic mean free path ($\lambda$) 
resulting in an enhanced probing depth. HAXPES has been successfully utilized in 
the study of Heusler TIs.\cite{ouardi:2011,shekhar:2012}
The SBB can be directly measured in photoelectron spectroscopy by controlling 
the escape depth in the photoemission process to track the depth dependence of 
core level energies. Such controlling can be achieved by changing the photon 
energy and consequently the inelastic mean free path, as demonstrated by Himpsel 
\emph{et. al}~\cite{Himpsel:1983} for low photon energy regime. The precision of 
this approach however depends on the energy distribution and the determination 
of the Fermi edges for different photon energies. Here, we propose the angular 
resolved HAXPES as an alternative to control the photoelectron escape depth, 
keeping constant all the experimental parameters: photon energy, incidence angle 
and probed region. This is possible thanks to the high energy wide-acceptance 
objective lens setup specially developed for this 
purpose~\cite{Matsuda:2004,Matsuda:2005}. The objective lens enlarge the 
effective acceptance allowing the measurement of the photoelectron angular 
distribution with a fixed incident angle. In this work, we measured the angular 
distribution of the photoemission of Bi$_2$Se$_3$ core levels, from where we 
observed the electric potential variations from bulk to surface.
\begin{figure}[ht]
\center{\includegraphics[width=1.0\columnwidth]{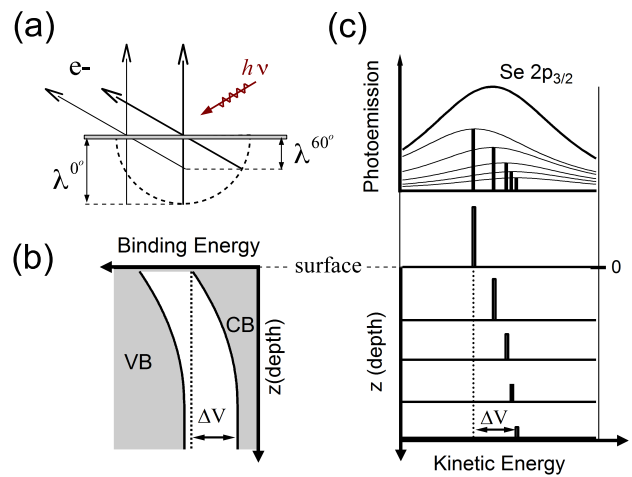}}
\vspace*{-0.0cm} \caption{\small  (a) Schematic of the experimental geometry. 
The angular distribution of photoemission
is simultaneously measured in the range from 0 (normal) to 60$^\circ$, being 
related with the electron escape depth $\lambda^{\theta}$, where $\theta$ represents
the emission angle.
(b)Model for SBB in Bi$_2$Se$_3$, where $\Delta V$ represents the bending 
amplitude. (c) Representation of normal photoemission of Se~2p$_{3/2}$. The 
vertical bars indicates the depth-dependent contribution of Se~2p$_{3/2}$  to the normal 
photoemission.}
\label{fig:fig_setup}
\end{figure}

Bi$_2$Se$_3$ single crystals were synthesized from stoichiometry mixture of high 
purity ($>99.99\%$) of bismuth (Bi) and selenium (Se).
The elements were sealed in a dry quartz ampoule under a pressure of 10$^{-6}$ 
Torr. The sealed ampoule was loaded into a vertical furnace
and heated to 800$^{\circ}$C at a rate of 60$^{\circ}$C/hour
followed by  12 hours soaking. For a single crystal growth, the temperature was 
slowly reduced from 800$^{\circ}$C to 500$^{\circ}$C and thereafter
by 100$^{\circ}$C/hour to room temperature. This procedure resulted in 
silver-colored single crystals size of 10 mm.

For HAXPES measurements, the crystal sample was exfoliated and kept in the air 
for few seconds in order to ensure the saturation of adsorption process
at surface, in such a way to eliminate any time-dependent effect. The experiment
was performed at BL47XU at Spring-8 (Japan) using 7.94
keV photon energy and $\pi$-linearly polarized light. The energy and angular 
distribution
of the photoexcited electrons was analyzed using a high energy VG Scienta 
R4000-HV
hemispherical analyzer. The objective lens, set in front of  the analyzer, 
enlarge the effective acceptance angle to about $60^{\circ}$ with an angular 
resolution of $1^{\circ}$. The homogeneity and
precision of the system was checked by mapping the angular distribution of Au 
$4f$ peaks.
The overall energy resolution was about 230~meV. The angle between the electron
spectrometer and the photon propagation was fixed at $90^{\circ}$.
Incoming photons was set to impinge on the sample at $60^\circ$ from its surface
normal, in such a way that the angular distribution of incoming electrons is 
measured from -2 to $62^\circ$ with respect to the sample normal. Sample temperature was 
kept at
40~K.  Figure~\ref{fig:fig_setup}a illustrates how the depth profile of core 
shift can be extracted
from the angular distribution of the photoemission by considering an effective 
escape depth $\lambda^\theta=\lambda \cos(\theta)$,  being $\theta$ the emission 
angle. Figure~\ref{fig:fig_setup}b and 1c show the relationship between SBB and 
the depth dependence of a core level energy shift.  The vertical bars in Figure 1c indicate the 
energy position of the Se $2p_{3/2}$ peak for different depths.
 $\Delta V$ represents the bending amplitude in both figures. The depth dependence of the energy shift 
of core levels mimics the band bending profile.

\begin{figure}[t]
\center{\includegraphics[width=0.80\columnwidth]{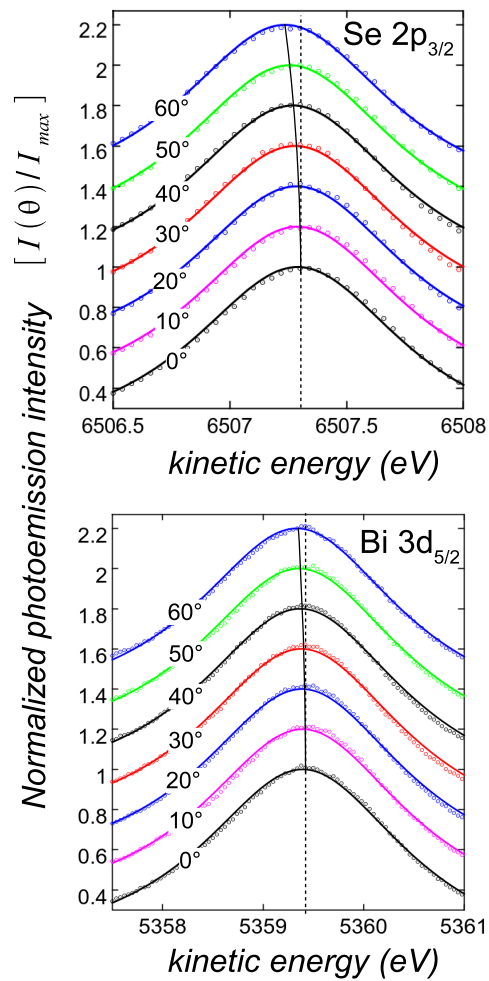}}
\vspace*{-0.0cm} \caption{\small (Color online): Energy distribution curves for  
Bi~3d$_{5/2}$ (top) and Se~2p$_{3/2}$.
Symbols and solid lines represent  experimental data and the calculated spectra 
$I(\theta,E_k)$ respectively.
The energy shifts are indicated by the vertical solid and dashed lines.
} \label{fig:fig_data}
\end{figure}

Symbols in Figure~\ref{fig:fig_data} represent the experimental energy 
distribution curves (EDC) for Se~2p$_{3/2}$ and Bi~3d$_{5/2}$  core state 
photoemission. The spectra are summed up over slices of $\pm2^{\circ}$ about the photoemission 
direction indicated by the label on the right of each curve. The curves were  normalized by the peak intensity.
Direct inspection of the EDC indicates an energy shift in the core level peaks,
as represented by the solid curved line. A total shift of -75(1)~meV from 
$\theta=0^{\circ}$ to $\theta=60^{\circ}$ for both Se~2p$_{3/2}$ and Bi~3d$_{5/2}$ peaks was determined by Voigt function fitting.
$\lambda$ is calculated by TPP-2M formula\cite{Tanuma:1994} to be  
$\lambda_{Bi}=\lambda(5.3keV)=8.7$nm and $\lambda_{Se}=\lambda(6.5keV)=9.4$nm for Bi$3d_{5/2}$ and 
Se$2p_{3/2}$ respectively. Providing the large electron escape depth and kinetic energies,
the measured EDC can only be originated by a bent 
potential which extends from surface up to a distance of the same order of magnitude than $\lambda$.
Therefore, the observed shifts indicate that bulk core-levels  ($z>z_l$) have smaller binding energy than core-levels near 
to the surface ($z\sim 0$).  This remarkable result consists in the direct observation 
of the band bending in  Bi$_2$Se$_3$, indicating a downward bending from bulk to surface.

For a quantitative analysis, we model 
the angular distribution $I(\theta,E_k)$ by the following equation:
\begin{equation}
    I(\theta,E_k)= \int_0^\infty dz ~ e^{-z/\lambda \cos(\theta)} \biggl[ 
\textsl{L}\bigl(\Gamma,b,E_0(z)\bigr) + I_o \biggr]
\end{equation}

where $I_0$ represents the constant background, and $L$ is a Lorenztian 
function. The Lorentzian shape is given by $b$ and $\Gamma$, representing the background 
asymmetry and  spectral linewidth respectively. The peak is centered in  different energies $E_0$
according to the depth position $z$.  $E_0(z)$ mimics
 the band bending (see Figure 1).


\begin{table}[h]
\caption{Optimized parameters for  
Bi~3d$_{5/2}$ and Se~2p$_{3/2}$ peaks. The variation of $E_0(z)$ from bulk value is shown in Figure 3.
}
\begin{tabular}{c  c  c}

Parameters&Bi~3d$_{5/2}$ &Se~2p$_{3/2}$ \\[0.5ex]

\hline
$  E_0^{bulk} $      &  5359.52~eV     &   6507.40~eV  \\
$\Gamma$   &  $2.47$~eV    &     $1.20$~eV    \\
$b$   &   $-0.40 \times 10^{-2}$    &     $-3.20 \times 10^{-2}$    \\

\end{tabular}

\end{table}

The band bending depth profile depends on detailed characteristics of the charge 
distribution near the surface. As simplest approach
we modeled the band bending profile as a quadratic function, extending 
for a distance $z=z_l$ from the surface ($z=0$). This approach allows the evaluation
of the extension $z_l$ and strength $\Delta V$ of the band bending in Bi$_2$Se$_3$ by refining $E_0(z)$:

\begin{equation}
  E_0(z) = \left\{
      \begin{array}{l l}
         \frac {\Delta V}{z_l}\left (z-z_l \right)^2 +  E_0^{bulk} & \quad z<z_l\\
        E_0^{bulk}                           & \quad z \geq z_l\\
      \end{array} \right.
\end{equation}

\begin{figure}[tb]
\center{\includegraphics[width=0.80\columnwidth]{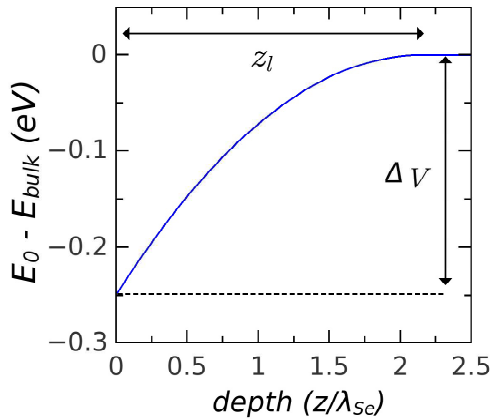}}
\vspace*{-0.0cm} \caption{\small (Color online): Energy shift depth profile for 
Se~$2p_{3/2}$ and Bi~$3d_{5/2}$.
The energy shift profile mimics the band bending in the Bi$_2$Se$_3$.}
\label{fig:fig_bb}
\end{figure}

For comparison with the experimental data, the calculated spectra 
$I(\theta,E_k)$ were convoluted with a Gaussian representing the total energy 
resolution (FWHM=$230~meV$). The curves were refined by the minimization of the mean
square error (MSE). The optimized parameters for Se~$2p_{3/2}$ and Bi~$3d_{5/2}$ 
peaks are summarized in Table I.  The optimum values for $z_l$ and $\Delta V$ 
are respectively 2.13$\lambda_{Se}$ (or 2.29$\lambda_{Bi}$)  and 0.25~eV.  
The calculated spectra $I(\theta,E_k)$ using  optimum parameters are shown in 
Figure~\ref{fig:fig_data} as solid lines.
Figure 3 shows the depth dependence  of the energy shift 
of both core levels, as derived from  $\Delta V(z)=[E_0(z)-E_0^{bulk}]$. 
The $z$ axis is shown normalized by the $\lambda_{Se}$.
The uncertainty in the estimation of  $E_0(z)$ coefficients was calculated  
by finding the ranges within the variation of the total MSE from the optimum value is smaller than 1\%.
It results in the regions:  $z_l \lambda_{Se} =$1.95--2.25 and 
$\Delta V=$0.23--0.26~eV. Therefore, our model suggests that the band banding
extends to   $z_l \thickapprox 2\lambda$ (about 20~nm).  These results are in agreement with  the band bending inferred 
from surface states position in  ARPES measurements~\cite{benia2011,King2011}.

In summary, we have observed the band bending Bi$_2$Se$_3$
by measuring the angular dependence of high energy 
photoelectrons emitted from core levels.
The angular dispersions in kinetic energy were modeled as originated from a 
depth-dependent shift in binding energy of core states.
The depth profile of core shifts was extracted in a single experiment from the 
angular distribution of photoelectrons.
 We found a downward band bending of about 0.25~eV, which extends to 
approximately 20~nm into the crystal,  in good agreement
with the values deduced from ARPES measurements~\cite{benia2011,King2011}.
Finally, it should be emphasized that the use of the angle-resolved HAXPES in 
wide angle allows the deeply probing of bulk states and opens the avenue
for investigation of band bending and interface potential in multilayered 
structures.

This work was financially supported by the Deutsche Forschungsgemeinschaft
(DFG, German Research Foundation) within the priority program SPP1666 "Topological insulators".
The synchrotron radiation measurements
were performed at BL-47XU with the approval of the
Japan Synchrotron Radiation Research Institute (JASRI)
(Proposal No. 2012A0043)

\bigskip



\end{document}